# A Cloud-Edge Framework for Energy-Efficient Event-Driven Control: An Integration of Online Supervised Learning, Spiking Neural Networks and Local Plasticity Rules

**Reza Ahmadvand, Sarah Safura Sharif, Yaser Mike Banad**
Department of Electrical and Computer Engineering, University of Oklahoma, Oklahoma, Unites States

**Email:**
iamrezaahmadvand1@ou.edu, s.sh@ou.edu, bana@ou.edu

**Abstract**
This paper presents a novel cloud-edge framework for addressing computational and energy constraints in complex control systems. Our approach centers around a learning-based controller using Spiking Neural Networks (SNN) on physical plants. By integrating a biologically plausible learning method with local plasticity rules, we harness the efficiency, scalability, and low latency of SNNs. This design replicates control signals from a cloud-based controller directly on the plant, reducing the need for constant plant-cloud communication. The plant updates weights only when errors surpass predefined thresholds, ensuring efficiency and robustness in various conditions. Applied to linear workbench systems and satellite rendezvous scenarios, including obstacle avoidance, our architecture dramatically lowers normalized tracking error by 96% with increased network size. The event-driven nature of SNNs minimizes energy consumption, utilizing only about $11.1 \times 10^4$ pJ (0.3% of conventional computing requirements). The results demonstrate the system's adjustment to changing work environments and its efficient use of computational and energy resources, with a moderate increase in energy consumption of 27.2% and 37% for static and dynamic obstacles, respectively, compared to non-obstacle scenarios.

Keywords: Spiking neural network, Cloud-edge framework, Obstacle Avoidance, Supervised learning, Local plasticity

## 1. Introduction

Autonomy in the new generation of robotic systems is dominated by the complexity of the task along with energy efficiency, reliability, and latency of the adopted frameworks. These challenges are increasingly being addressed with advancements in brain-inspired computing, such as neuromorphic systems [1]. Particularly, for dynamical systems leveraging centralized control methods, where the aforementioned problems limit the implantation of the control systems, cloud-based control systems are proposed to relax such limitations [2, 3]. In cloud-based systems, the physical location of the controller that adopts the control frameworks and algorithms has been moved from the physical plant to the cloud. However, these systems introduce new challenges, such as the need for continuous communication between the physical plant and the cloud, leading to potential signal access issues in remote locations along with high energy consumption for data transfer. Thus, having an energy-efficient, fast, and reliable learning-based controller on the plant, where the generated data can be processed, becomes paramount. Spiking neural networks (SNNs), which emulate brain-like computation, emerge as an energy-efficient alternative with low latency for complex data processing and decision-making on the plant [4].

From the other perspective, the effectiveness of these networks largely depends on their ability to continuously adapt to evolving learning rules. Particularly, for on-chip implementation, these learning rules must adhere to three principles: (1) adopting event-based computation through discrete spikes to guarantee superior energy efficiency; (2) facilitating local parallel information processing for cost-effective, near-memory, and asynchronous computation; and (3) enabling low-latency inferencing in neurons without the need for dependency on future data [5]. Generally, the training rules proposed for SNNs can be classified into three categories: shadow training, gradient-based methods, and local learning rules. Shadow training suffers from high energy consumption and latency issues. Gradient-based methods struggle with biological implausibility and need for non-local information. This study focuses on local learning rules, which use only available local information for synaptic updates, closely mirroring biological systems [6].

In biologically plausible learning rules, which is the concern of this study, the synaptic weight updates are just dependent on the information that is spatially and temporally local to the desired synapse without dependency on the global error of the network [5, 6]. For instance, spike-timing-dependent-plasticity (STDP) and its variants such as reward-modulated STDP (R-STDP) which draws inspiration from dopamine-driven learning in the biological nervous system, anti-STDP, and remote supervised method (ReSuMe). These biologically plausible local learning rules function based on the temporal correlations between spike timings of pre- and post-synaptic neurons [7, 8, 9].

The application of these local learning rules in SNNs has been instrumental in harnessing the benefits of neuromorphic structures within robotic systems such as enhanced energy efficiency, scalability, and robustness against neuron silencing. Leveraging supervised local learning rules, SNNs have been implemented for various applications [1, 10]. For instance, an adaptive control has been introduced [11] for a three-link robotic arm based on local spike-based learning rules. In another work [12], a spiking neural network has been trained using the STDP approach to control a four-degree-of-freedom robotic arm, implementing a humanoid robot. Despite the diversity and potential of these local learning rules, they often fall short in simultaneously addressing the critical factors of locality, robustness, and spiking efficiency – key aspects for their practical deployment in neuromorphic hardware. To address these challenges, Alireza Alemi et. al. [13] have proposed a supervised local learning rule for learning arbitrary dynamics in nonlinear systems which can simultaneously meet the factors of locality, robustness, and spiking efficiency. The efficiency of spiking activities means that the desired method functions with the smallest possible number of spikes, ensuring minimal energy consumption due to emitting spikes. The approach integrates the principles of efficient-balanced networks (EBN) [14] with nonlinear adaptive control theories and considers the nonlinear dendritic computations. In this context, basis functions from adaptive control theories are leveraged to accommodate nonlinear dendritic computations, enabling learning of coefficients by exploiting error signal correlations. The proposed learning rule benefits from EBNs' spiking efficiency and robustness, incorporating the locality advantages of adaptive control theories.

Therefore, from this perspective and considering the recent advances in applications of neuromorphic computing and SNNs for the estimation and control of dynamical systems [15, 16, 17], our study aims to harness the advantages of these developments [13]. Drawing inspiration from supervised training methods applied to ANNs for control systems such as the cartpole [18], we propose a cloud-edge-based, online supervised training framework by employing a recurrent EBN composed of leaky integrate-and-fire (LIF) neurons. The key feature of our SNN is its ability to quickly learn the control policy, requiring only a limited number of epochs to replicate the control signal generated by a model-based controller operating on the cloud. A notable aspect of our approach is that the implemented SNN is designed to be independent of the underlying dynamics of the system and the controller. As such, none of the SNN parameters is required to be trained about the adopted controller (source of the desired signal) or dynamical plant (which receives the control signal reproduced by SNN at the edge). This independence makes our method highly versatile and applicable to a wide range of systems, regardless of the complexity of their dynamics.

To assess the efficacy of our approach, we first applied it to a linear workbench problem in which the SNN successfully learned to reproduce the control signal from a linear quadratic regulator (LQR) [19]. The SNN then used this learned signal to control a desired plant by reproducing this signal. Building on this, we extended our approach to a more complex and realistic scenario: satellite rendezvous with obstacle avoidance maneuver, considering both static and dynamic obstacles which is a crucial real-world problem in space robotics [20] in the presence of uncontrolled space objects/debris [21]. For this application, the model-based LQR controller on the cloud was designed using the Clohessy-Wiltshire (CW) equations [22]. Our findings revealed that the proposed framework offers acceptable performance in terms of accuracy and energy efficiency, surpassing the traditional frameworks in computational efficiency. Also, we have shown that the implemented SNN adjusts its spiking activity with changes in working conditions along with demonstrating that the level of energy consumption of SNN-based frameworks increases with the complexity of the considered task.

This paper is organized as follows: Section 2 provides an overview of related works and their contributions to the field. Then, the preliminaries, underlying theories, and the proposed framework for addressing the problem of concurrent robust estimation and control in linear dynamical systems are presented in Section 3. Section 4 presents numerical simulations and discussions of the results obtained from these simulations. Finally, Section 5 concludes the paper, summarizing key findings and implications.

## 2. Related works and Contributions

This section offers a comprehensive overview of recent advancements in the field and our distinctive contributions:

*2.1 Related works*

The recent literature has focused on the applications of SNN-based frameworks in dynamic systems. A notable example is an SNN-based framework that has been introduced for concurrent estimation and control in linear dynamical systems, especially applied to satellite rendezvous maneuvers [17]. This approach, an extension of the method by Slijkhuis *et al.* combining Luenberger observer and LQR controller [15], introduces the SNN-based Modified Sliding Innovation Filter (SNN-MSIF). The SNN-MSIF is a robust neuromorphic estimator for linear systems [16], leveraging the gain formulation introduced in [23]. Although these frameworks exhibit commendable accuracy, robustness, and reliability in the presence of modeling uncertainties, neuron silencing, and measurement outliers, they lack the implementation of any learning rule.

In contrast, in [11], a biologically plausible spiking neural model is proposed for adaptive control of a nonlinear three-link arm, adaptable to unknown changes in the environment and arm dynamics. This model draws inspiration from the localized neural computations of the motor cortices and cerebellum. Furthermore, in [12], an STDP learning rule-based SNN framework, autonomously learning to control a 4-DOF robotic arm, using the Izhikevic model for more biologically plausible behavior. The SNN framework has also been successfully implemented on the kinematics model of the arm of an iCub humanoid robot. While these studies show significant progress, they fall short in achieving optimal spiking efficiency, an area our research aims to address [13].

Addressing this gap, Alireza Alemi *et. al.* [13] developed a local supervised learning rule for SNNs, meeting locality, robustness, spiking efficiency, and versatility to be used for any arbitrary nonlinear dynamical system. This rule, tested on the Lorenz system, has shown promising results in supervised learning of dynamic systems but has not been extensively applied to control problems. Therefore, the presented study has focused on introducing a cloud-edge framework using supervised learning for dynamical physical plants [13].

*2.2 Contributions*

Our research contributions are multifaceted:

- **Development of Cloud-Edge Framework:** Building on the learning rule in [13], we propose an online biologically plausible supervised learning approach using local plasticity rules leveraging a cloud-edge framework. This innovative strategy represents an advancement in SNN-based control systems.
- **Application to Satellite Rendezvous Problem:** Our implementation of the proposed rule showcases its applications in space robotics, particularly in satellite rendezvous scenarios. This represents a novel application of learning-based neuromorphic control strategies.
- **Energy Efficiency and Accuracy Assessment in Resource-constrained Control Systems:** Our research delves into the balance between energy efficiency and accuracy, highlighting how the complexity of tasks impacts the energy consumption of SNNs and underscoring the advantages of event-driven control systems.

## 3. Theory

This section elucidates the theoretical underpinnings and outcomes of our study, focusing on the application of Spiking Neural Networks (SNNs) in centralized control systems. Our approach leverages cloud-based computing to address computational complexities in centralized control systems [2, 3]. In this cloud-centric architecture, control signals are computed in the cloud and transmitted to the physical plant via various antennae. While this setup necessitates continuous plant-cloud communication, it also poses a risk of losing control in case of communication failures or interruptions. To address this challenge, our study introduces an innovative architecture, as presented in Figure 1. This architecture features an SNN-based onboard controller on the plant, capable of online training to reproduce control signals generated in the cloud. This design significantly relaxes the need for continuous plant-cloud communication, thereby enhancing the system's reliability.

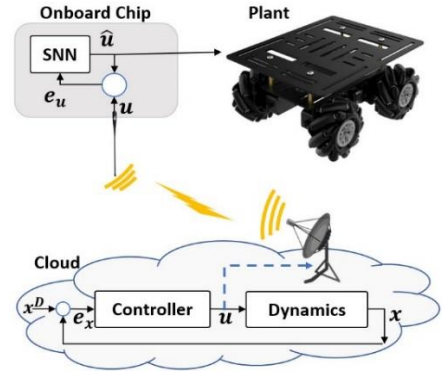

**Figure 1**. A schematic representation of a cloud-edge framework employing online supervised learning. The 'Rover' serves as an illustrative example for plant within an edge environment.

Moreover, our proposed architecture in this study leverages the advantages of neuromorphic computing such as energy efficiency and scalability inherent in SNNs, compared to traditional computing frameworks.

*3.1 SNN and local learning rule*

A review of the method and its foundational theories is presented in applying the SNN architecture and learning rule [13]. SNNs, as brain-mimicking computational models, are biologically plausible and effectively emulate neural circuitry in the human brain, in which the neurons communicate via spikes. Given the prevalent use of Leaky Integrate-and-Fire (LIF) neurons in robotics and engineering [5], our focus here is on developing a recurrent EBN comprising these neurons from the perspective of EBN and spike coding theory. , To develop an SNN capable of representing temporal variation of parameters such as $x$, two key assumptions are necessary [13, 15, 14]. First, as it has been expressed in Eq. (1), using a linear decoder matrix $D$, it is needed for $x$ to be extracted from the filtered spike trains.

$$\hat{x} = Dr \qquad (1)$$

In this equation, $r \in R^N$ represents the filtered spike trains, which are the convolution of spike trains with synaptic kernels, and can be interpreted as instantaneous firing rate [13], and exhibit slower dynamics compared to $s \in R^N$. The dynamics of these filtered spike trains obeys the following expression:

$$\dot{r} = -\lambda r + s \qquad (2)$$

The second assumption is that the network minimizes the cumulative error between the actual value of $x$ and the estimated $\hat{x}$, focusing on optimizing spike timing rather than changing the output kernel values $D$. So, the network minimizes the cumulative error between the state and its estimate while limiting computational cost by controlling spike occurrence. To achieve this, it minimizes the following cost function [13, 14]:

$$J = \frac{1}{t}\int_0^t (\|x(\tau) - \hat{x}(\tau)\|_2^2 + \nu\|r(\tau)\|_1 \\ + \mu\|r(\tau)\|_2^2 \ ) d\tau \qquad (3)$$

Here, $\|.\|_2^2$ represents the Euclidean norm, and $\|.\|_1$ indicates L1 norms. This firing rule ensures that each neuron emits a spike only when it contributes to reducing the predicted error, a concept akin to predictive coding [14]. The network's neurons have membrane potentials and firing thresholds, described by Eq. (4) and Eq. (5):

$$\sigma_i(t) = D_i^T(x(t) - \hat{x}(t)) - \mu r_i \qquad (4)$$
$$T_i = (D_i^T D_i + \nu + \mu)/2 \qquad (5)$$

Here, $\sigma_i \in R$ denotes the membrane potential of $i^{th}$ neuron. $D_i$ is the $i^{th}$ column of the matrix $D$, reflecting the neuron's output kernel and the change in the error due to a spike of $i^{th}$ neuron. The parameters $\nu$ and $\mu$ control the trade-off between computational cost and accuracy, with $\nu$ encouraging the network to use as few spikes as possible, while the quadratic term is influenced by $\mu$, which is responsible for a uniform distribution of spikes among network neurons. Proper tuning of these parameters yields biologically plausible spiking patterns where neural activity is distributed approximately evenly among neurons. This firing rule operates based on the concept that the network minimizes the cost function by controlling spike occurrence in response to excitation and inhibition, ultimately, reducing the prediction error [13, 14]. After differentiating and simplification of Eq. (4), the dynamics of membrane potential for a SNN that can mimic any arbitrary dynamics like $\dot{x} = f(x) + c(t)$ has been achieved [13]:

$$\dot{\sigma} = -\lambda\sigma + D^T c - (D^T D + \mu I)s \\ + D^T(\lambda x + f(x)) \qquad (6)$$

The above expression shows that the optimal weight matrix for the fast connections is $\Omega_f = D^T D + \mu I$ and also the optimal encoding matrix is $D^T$. As it is stated in [13], for being able to implement any arbitrary dynamics in this framework, two approximations are made. Firstly, inspired by theories of adaptive nonlinear control, the last term $\lambda x + f(x)$ is approximated by a weighted sum of basis functions such as $\lambda x + f(x) = \sum_i \Omega_i \phi_i(x)$, where $\Omega_i$ is $i^{th}$ column of $\Omega$ and $\phi_i(.)$ can be any sigmoidal nonlinearity that here is set to tanh( ). Second, based on Eq. (1), the state $x$ is replaced by its estimate $Dr$, culminating in a refined expression for the dynamics of membrane potential, as follows:

$$\dot{\sigma} = -\lambda\sigma + D^T c - \Omega_f s + D^T \Omega^T \phi(Dr) \qquad (7)$$

Biologically, in the above expression, the last term corresponds to the nonlinear and highly structured dendrites that perform nonlinear operations in the biological brains. So, to relax the constraints that are caused by the fact that, while the dendrites are often nonlinear their inputs are stochastic, the term $\phi(Dr)$ is replaced by $\psi_i(r) = \phi_i(M_i^T r + \theta_i)$ which is a highly unstructured nonlinear dendrite that can receive stochastic inputs from other neurons. where $M$ is the random fixed matrix, and $\theta_i$ is random noise. This substitution is justified by the fact that the internal dynamics of the network are highly robust to great amounts of noise when the inequality $N > 2K$ is met [13]. The network dynamics are thus defined by the following expression:

$$\dot{\sigma} = -\lambda\sigma + Fc - \Omega_f s + \Omega_s \psi(r) \qquad (8)$$

Here, $\lambda$ represents the leak rate for the membrane potential, and $F = D^T$ encodes the transformation of dynamics external stimulation into a spike sequence interpretable by the network. The synaptic weights, $\Omega_s$ and $\Omega_f$ correspond to slow and fast connections, respectively. Slow connections typically govern the implementation of the desired system dynamics, whereas fast connections are crucial for uniformly distributing spikes across the network [13, 17, 14]. To incorporate a local learning rule based on adaptive control theories, the error $e(t) = x(t) - \hat{x}(t)$ is not fed back into the whole network via the input $c(t)$. Instead, the error is directly injected into each neuron using the optimal encoder $D^T$. Consequently, with such neuron-specific error feedbacks, the network equation is reformulated as:

$$\dot{\boldsymbol{\sigma}} = -\lambda\boldsymbol{\sigma} + F\boldsymbol{c} - \Omega_f \boldsymbol{s} + \Omega_s \psi(\boldsymbol{r}) + kD^T \boldsymbol{e} \quad (9)$$

As indicated earlier, $\Omega_s$ plays a crucial role in executing the desired dynamics, necessitating its learning from example trajectories of an external unknown dynamic in a teacher-student fashion. To this aim, the following expression from [13] outlines this learning process:

$$\dot{\Omega}_s = \eta \psi(\boldsymbol{r})(D^T \boldsymbol{e})^T \quad (10)$$

This expression demonstrates that the proposed weight update relies on pre-synaptic inputs, which will be provided via dendritic nonlinearities, and the projection of network error both from the last time step of implementation. Therefore, it can be inferred that the learning rule is both spatially and temporally localized [7, 13].

### 3.2 Online supervised learning architecture

As depicted in Figure 2, the architecture for online supervised learning of SNNs utilizes a specific learning rule where the SNN, positioned on the plant, receives a control signal from cloud-based algorithms. The SNN's objective is to learn and replicate this signal until the error $\boldsymbol{e}_u = \boldsymbol{u} - \hat{\boldsymbol{u}}$ drops below a predefined threshold $\boldsymbol{e}_{th}$. Once this threshold is reached, the SNN continues to control the plant independently, without further weight updates. Here, $\boldsymbol{u}$ represents the cloud-provided control signal, and $\hat{\boldsymbol{u}}$ denotes the SNN-reproduced signal. It is notable that, in the considered strategy, the SNN-based edge controller maintains communication with the cloud for the initial 50-time steps. Subsequently, it reduces communication with the cloud for all the time steps, and only checks the threshold crossing every 50-time steps.

A key aspect of this approach is the SNN's independence from both the source of the control signal and the target plant. This means the SNN requires no training about the adopted controller on the cloud or the dynamical plant, enabling easy implementation on various plants with different control methods.

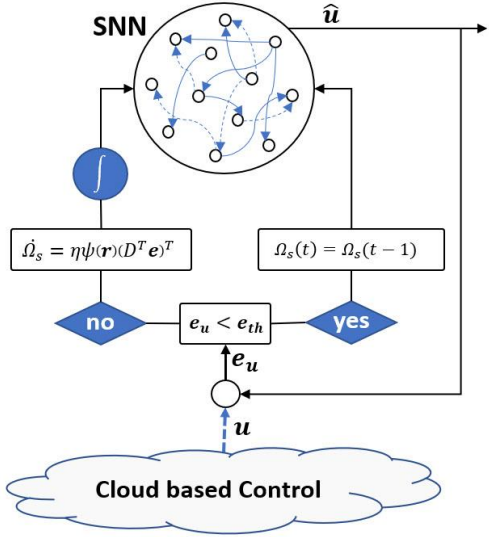

**Figure 2**. Block diagram of the constrained weight update rule

## 4 Numerical Simulations

In this section, we first apply the proposed framework to a linear workbench problem, evaluating its performance in terms of accuracy and energy efficiency. We then extend the analysis to a real-world scenario, focusing on the optimal control of satellite rendezvous maneuvers.

### 4.1 Linear workbench system

The performance investigation begins with applying the proposed strategy to the following linear system:

$$\dot{\boldsymbol{x}} = \begin{bmatrix} 0 & 1 \\ -2 & 0 \end{bmatrix} \boldsymbol{x} + \begin{bmatrix} 0 \\ 1 \end{bmatrix} \boldsymbol{u} \quad (11)$$

where:

$$\boldsymbol{u} = -K_c \boldsymbol{x} \quad (12)$$

In this system, $K_c$ represents the controller gain, designed using optimal control theories as detailed in the next section. Simulations ran for a 10-second duration with a 0.1-second timestep, employing parameters from Table 1. The decoding matrix $D$'s elements were sampled from a zero-mean Gaussian distribution with a covariance of 10.

**Table 1**. Linear system simulation parameters

| Parameter | Value |
|---|---|
| $\boldsymbol{x}_0$ | $[5,2]^T$ |
| $Q_c$ | $(1e-6)I_6$ |
| $R_c$ | $I_3$ |
| $K_c$ | $[0.2361, 1.2133]^T$ |
| $N$ | 30 |
| $\boldsymbol{e}_{th}$ $(N)$ | $(1,1)/10$ |
| $\lambda$ | 0.001 |
| $\mu$ | 0.001 |
| $\nu$ | 0.001 |
| $k$ | 500 |
| $\eta$ | 0.001 |

Figure 3 presents the controlled states obtained from the physical plant under the SNN-reproduced signal $\hat{\boldsymbol{u}}$ for $N = 5$, 15 and 30, compared with states from the cloud's dynamical model using the LQR-provided control signal $\boldsymbol{u}$. Figure 3(a) reveals that the obtained results for the state $x_1$ of the physical plant using $\hat{\boldsymbol{u}}$ have different accuracies for various $N$. The result for $N = 5$ exhibits some deviation from the cloud-provided state, indicating that larger $N$ values enhance tracking accuracy. Increasing the $N$ has increased the tracking accuracy in a way that for $N = 30$, the result from the physical plant almost aligns with the dynamic model on the cloud, considering the expected values for $x_1$. Thus, it can be deduced that the network size influences controller accuracy. Also, the physical plant satisfactorily follows the expected state trajectory using $\hat{\boldsymbol{u}}$. Figure 3(b) demonstrates similar observations for the state $x_2$. The zoomed inset highlights deviations from the desired values, confirming a 5 times higher deviation for $N = 5$ in the considered area, compared to $N = 30$ for both the states $x_1$ and $x_2$. These observations underscore the impact of network size on controller accuracy and the effective tracking of desired state trajectories.

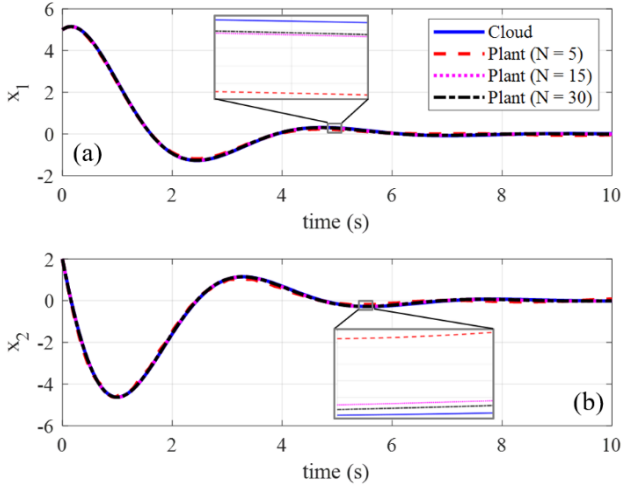

**Figure 3.** Time history of controlled states obtained from the plant for $N = 5, 15$ and $30$ compared with expected states provided by cloud for (a) state $x_1$ and (b) state $x_2$.

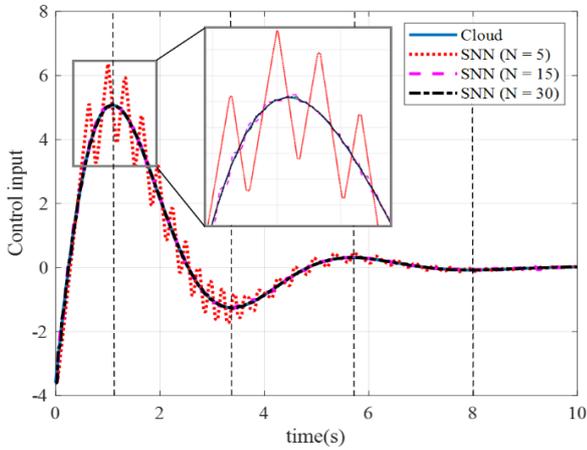

**Figure 4.** Time history of control input reproduced by SNN on the plant for $N = 5, 15$ and $30$ compared with expected input provided by the cloud. Notice, that vertical dashed lines specify where the behavior of the desired signal changes.

Figure 4 compares the control input $\hat{u}$ reproduced by the SNN with the original input $u$ provided by the cloud, albeit with noticeable fluctuations, which vary depending on neuron counts, $N$. These fluctuations are more pronounced in regions where the desired signal exhibits higher nonlinearity. Specifically, with a neuron count $N = 5$, the fluctuation domain is significantly larger between 0.5 and 4 seconds compared to the behavior of SNN tracking signals after t = 8s, where it significantly dampens. The inset shows the region where the signal changes the rising to declining trends, corresponding to the most nonlinear region of the curve. It becomes clear that an increase in neuron count $N$ leads to improved tracking performance by the SNN, characterized by reduced fluctuation domains in these critical regions. For instance, in the magnified area, the fluctuation domain at $N = 5$ is about 10.5 times larger than at $N = 15$, and nearly triple that of $N = 30$. This observation suggests that a higher neuron count is required to respond to any abrupt maneuver, enhancing the accuracy of the SNN's signal reproduction relative to the cloud-provided signal, and consequently improving the control performance of the system. For clarity, Figure 4 only displays the results for $N = 5, 15$ and $30$.

Further investigation into the effect of neuron count on accuracy is presented in Figure 5. In Figure 5(b), the variation of normalized tracking error reconfirms that an increase in $N$ leads to enhanced SNN performance, with a notable normalized error reduction of 96%. However, beyond $N = 30$ (Region 2 in the figure), the increase in neuron count does not significantly affect accuracy, which remains nearly constant for the remaining simulation cases. This behavior reflects the intrinsic scalability of SNNs. This means that they can maintain a redundant resource of spiking neurons that ideally do not consume energy when not in use, remaining in standby mode and ready to engage in data processing for more complex conditions as needed. Thus, for the conventional design, it is better to consider more neurons than the $N$ which corresponds to minimal error to maintain neuron silencing in the neuromorphic system for the conditions in which the SNN encounters more complex problems.

Figure 5(a) demonstrates that the SNN, in accordance with the EBN theory, operates with minimal spike usage. In this problem, the SNN utilized 1444 spikes during the simulation period, which is only 4.81% of the potential 30,000 spikes (for 30 neurons across 1,000 time steps). With an energy consumption of 23.6 pJ per spike in neuromorphic hardware (e.g., intel's Loihi neuromorphic chip) [24], the SNN would consume only 3.4078×104 pJ of energy for the task considered here and this is much more efficient in terms of energy consumption.

Moreover, it is noticeable that each vertical dashed line on the spiking pattern corresponds to the vertical dashed lines exhibited in Figure 4. It is clearly apparent that when the graph of desired input presented in Figure 4 passes from the vertical dashed lines, the changes in the direction of desired values cause the change in the spiking activity regime between the vertical dashed lines demonstrated in Figure 5(a). Thus, it can be deduced that corresponding to any change in the desired value, the SNN adjusts its neural activity.

Figure 5(b) and 5(c) demonstrate the changes in the tracking accuracy of the SNN and its energy consumption versus the neuron counts respectively. The obtained results, reconfirm that the reduction of neuron counts will lead to declination in energy consumption of the SNN. They demonstrate that for the considered system in this part, we can reduce the neuron counts from 50 to 30 to decline the energy consumption without experiencing significant change in the tracking accuracy in region 2 of Figure 5(b), but the further reduction of neuron counts below 30 in Region 1 of Figure 5(b) will lead to energy efficiency at the cost of network tracking accuracy.

Figure 6 illustrates the weight update of a single neuron (elements of the 1st row of $\Omega_s$) for two different cases of $N = 10$ and $N = 50$ over the simulation time period. In the presented graphs in Figure 6 each line corresponds to the temporal variations of a synaptic weight. Figure 6(a) and Figure 6(b) reveal the temporal variations for the cases of $N = 10$ and $N = 50$ respectively. The depicted graphs confirm that, owing to the update rule presented in Figure 2, for both cases the weight update rule wasn't implemented continuously over the simulation time. In contrast to what it has been shown for the case of $N = 50$ which its weights have immediately converged to a constant set and it continues without any weight update until the end of the simulation, for the case of $N = 10$ the update rule is implemented continuously for the first 50 time steps (until $t = 0.5$s), then it continued with constant weights until almost $t = 1.8$s which it has updated its weights that demonstrate the proposed framework encountered to a tracking error threshold crossing. Furthermore, a comparison between these figures demonstrates that, in comparison with the obtained results for the case of $N = 10$, although the synaptic weights have converged to constant values in almost $t = 0.5$s, for the case of $N = 50$ the synaptic weights exhibited the faster convergence to constant values I which they have converged in a time less than $t = 0.1$s. therefore, it can be inferred that, increasing the number of neurons will increase the rate of convergence of synaptic weights. Moreover, the depicted graphs exhibited almost symmetrical patterns with respect to zero which can reconfirm that in the EBN the level of excitation of neurons is almost equal to the level of inhibition [13].

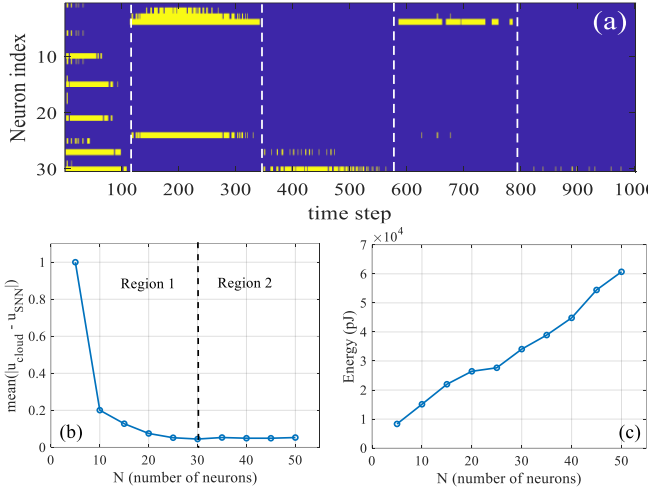

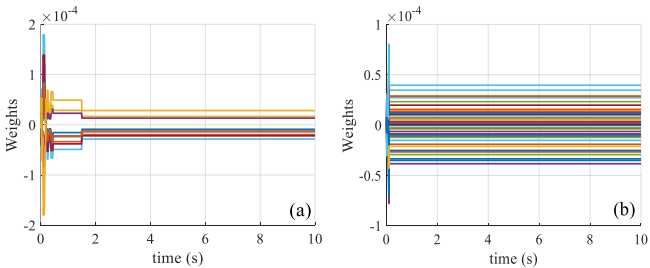

**Figure 5**. (a) Sequence of spikes versus time for 30 neurons, Vertical dashed lines specify where the spiking pattern regime changes (b) Normalized tracking error versus $N$, (c) Energy consumption vs. $N$.

**Figure 6**. Temporal variation of synaptic weights of a single neuron (elements of the 1st row of $\Omega_s$) for the case of (a) $N = 10$ and (b) $N = 50$.

Conclusively, the results in this section have demonstrated the efficiency of the proposed framework. Notably, it excels in accuracy and significantly mitigates the need for continuous plant-cloud communication during all operational steps. This reduction in data transfer between the plant and cloud substantially lowers energy consumption, a crucial benefit in today's energy-constrained systems. Additionally, the findings underscore that, unlike conventional methods, the integration of neuromorphic architectures, owing to the scalability and efficiency of SNNs, facilitates more reliable and energy-efficient control systems on the plant. Moreover, the versatility of our proposed framework allows for its easy adaptation to centralized control scenarios in dynamical systems manipulation.

### 4.2 Satellite rendezvous without obstacle

In this section, the proposed framework has been applied to the problem of satellite rendezvous as one of the fundamental challenges for various space missions, including assembling, servicing, or refueling spacecraft in orbit [20]. As it has been shown in Figure 7, the satellite rendezvous is a multi-object maneuver between two satellites in which the chaser satellite approaches the target and reaches together on the rendezvous point.

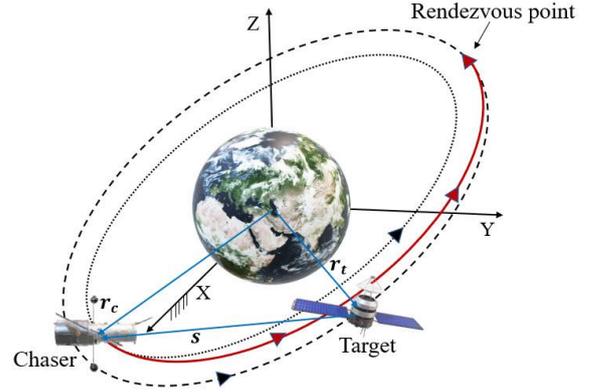

**Figure 7**. Schematic of satellite rendezvous problem. By permission from Ref. [17]

This study aims to implement the LQR controller on the cloud, utilizing Clohessy-Wiltshire (CW) equations. First, we will introduce the state space model for the satellite rendezvous problem to facilitate the development of the LQR controller. Following this, we will present the simulation results and discussions to demonstrate the effectiveness of the implemented controller.

The following system of differential equations represents the linearized Clohessy-Wiltshire equations which describe the relative motion in the target frame [22]:

$$\ddot{x} - 2n\dot{z} = f_x \quad (13)$$
$$\ddot{y} + n^2\dot{y} = f_y \quad (14)$$
$$\ddot{z} + 2n\dot{x} - 2n^2\dot{z} = f_z \quad (15)$$

Here, $n = \sqrt{\mu_{earth}/R_o^3}$, where $\mu_{earth}$ is the earth's gravitational parameter, and $R_o$ represents the orbital radius of the target spacecraft, and $n$ is the mean motion. Considering the state vector as $\boldsymbol{x} = [x, y, z, \dot{x}, \dot{y}, \dot{z}]^T$, an input vector $\boldsymbol{u} = [f_x, f_y, f_z]$, the state space form of the CW equations is:

$$\dot{\boldsymbol{x}} = A\boldsymbol{x} + B\boldsymbol{u} \quad (16)$$

where matrices A and B are given by the following equations:

$$A = \begin{bmatrix} 0 & 0 & 0 & 1 & 0 & 0 \\ 0 & 0 & 0 & 0 & 1 & 0 \\ 0 & 0 & 0 & 0 & 0 & 1 \\ 0 & 0 & 0 & 0 & 0 & 2n \\ 0 & 0 & 0 & 0 & -n^2 & 0 \\ 0 & 0 & 0 & -2n & 0 & 2n^2 \end{bmatrix};$$

$$B = \begin{bmatrix} 0 & 0 & 0 \\ 0 & 0 & 0 \\ 0 & 0 & 0 \\ 1 & 0 & 0 \\ 0 & 1 & 0 \\ 0 & 0 & 1 \end{bmatrix} \quad (17)$$

Generally, for any pair $(A, B)$ which met the controllability condition, the LQR control input can be computed using the expression in Eq (16), in which the $K_{LQR}$ needs to be designed to minimize the cost function, $J_c$, in Eq. (17) [19]:

$$\boldsymbol{u} = -K_{LQR}\boldsymbol{x} \quad (18)$$

$$J_c = \int_0^\infty (\boldsymbol{x}^T Q_c \boldsymbol{x} + \boldsymbol{u}^T R_c \boldsymbol{u})dt \quad (19)$$

In the above expression, $Q_c$ and $R_c$ are the weight matrices that can be determined through trial and error satisfying the conditions $Q_c > 0$ and $R_c \geq 0$. Finally, the LQR controller gain can be computed using $K_{LQR} = R_c^{-1}B^T S$, where $S$ is the unique positive semidefinite solution of the algebraic Riccati equation:

$$A^T S + SA - SBR_c^{-1}B^T S + Q_c = 0 \quad (20)$$

After designing the controller gain, we integrated the cloud-based control system for satellite rendezvous as shown in Figure 1. This integration utilized the strategy outlined in Figure 3, employing an SNN with $N$ neurons in the proposed cloud-based online supervised learning architecture. In this scenario, the chaser satellite acts as the desired plant, receiving the control signals generated by the SNN.

To validate our approach, we conducted simulations, based on parameters from Table 2, over a time duration of 360 seconds with a time step of 0.1 seconds.

It's important to note that in these simulations, the elements of decoding matrix $D$ are sampled from a zero-mean Gaussian distribution with a covariance of 1/1000. Figure 8 reveals the trajectories for the case of satellite rendezvous without obstacles. The comparison between the obtained trajectory from the plant with the trajectory expected from the cloud confirms that the trained SNN could satisfactorily perform the considered task. Figure 9 demonstrates the proposed framework, and how effectively it controls the states of the chaser satellite. The satellite's actual performance, guided by the SNN-generated control inputs, closely followed the expected state values calculated on the cloud.

**Table 2**. Paremeters for satellite rendezvous

| Parameter | Value |
| --- | --- |
| $\boldsymbol{r}_0$ (m) | $[70, 30, -5]^T$ |
| $\boldsymbol{v}_0$ (m/s) | $[-1.7, -0.9, 0.25]^T$ |
| $\boldsymbol{x}_0$ | $[\boldsymbol{r}_0, \boldsymbol{v}_0]^T$ |
| $Q_c$ | $(1e - 6)I_6$ |
| $R_c$ | $I_3$ |
| $\mu_{earth}$ $(km^3/s^2)$ | 398600 |
| $R_o$ (km) | 6771 |
| N | 50 |
| $\boldsymbol{e}_{th}$ (N) | (1,1,1)/10000 |
| $\lambda$ | 0.0001 |
| $\mu$ | 0.0001 |
| $\nu$ | 0.0001 |
| $k$ | 250 |
| $\eta$ | 0.001 |

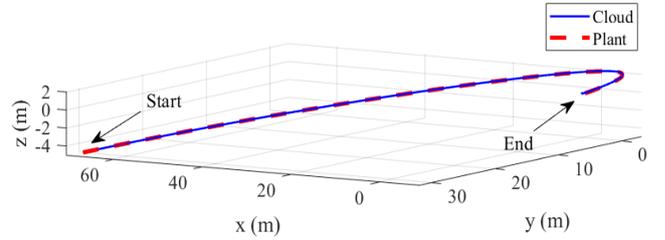

**Figure 8**. Comparison of obtained trajectory in 3D space for the plant under the SNN control with the expected trajectory provided by cloud for satellite rendezvous without obstacle.

Note: The supplementary video provides a visual demonstration of the satellite's navigation with respect to spiking activities and energy consumption in Figure11.

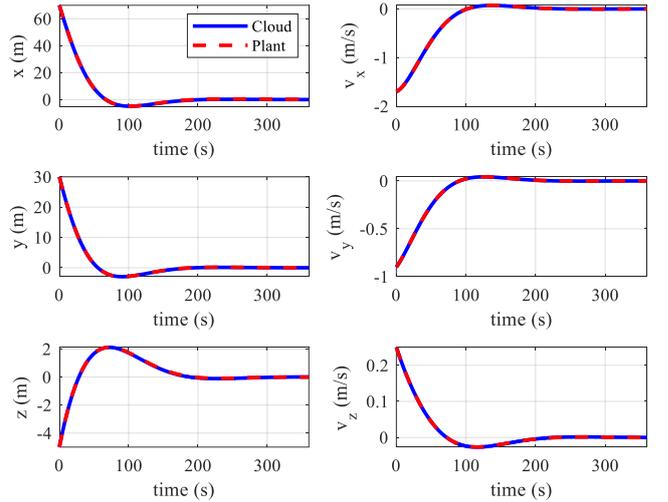

**Figure 9**. Time histories of obtained results for the controlled states from the plant (red dashed line) in comparison with cloud (blue solid line) for satellite rendezvous without obstacle.

Figures 10(a), 10(b) and 10(c) illustrate the results for the control input $\hat{u}$ elements reproduced by the SNN on the plant, compared with input $u$ provided by the cloud. These figures show that the SNN closely tracked the cloud-provided signal across all input vector elements, suggesting effective control of the physical plant. Figures 10(d), 10(e) and 10(f) show tracking errors for each input element for the first 5 sec. This early time frame is highlighted for a clear visual representation of error convergence. Notably, the errors converge to nearly zero around 2 sec and the system maintains minimal fluctuations thereafter. This rapid error convergence suggests the SNN's high responsiveness and precision in mimicking the control inputs, underlining its suitability for dynamic and time-sensitive applications like satellite rendezvous.

Figure 11(a) shows the spiking pattern of the SNN for the given task, demonstrating efficient task execution with minimal spike usage. The spiking pattern indicates three distinct spiking regimes corresponding to changes in control input behaviors: a shift in input direction around $t = 40s$ and input convergence to near zero at $t = 200s$. The SNN emitted 3434 spikes across 50 LIF neurons, just 0.19% of the potential 1,800,000 spikes, showcasing a biologically plausible spiking distribution in the network. Figure 10(b) shows the temporal variation of the consumed energy per time step for spike emission. It reveals that most of the energy is consumed before $t = 150s$ where the elements of the input vector are significantly greater than zero and after $t = 150s$, where the input vector elements are almost near to zero the consumed energy is considerably reduced, and this is vividly apparent in the obtained spiking pattern.

Figure 11(c) demonstrates the energy consumption versus time for performing the considered task. It exhibits that the energy consumption has drastically increased to $8.1 \times 10^4$ pJ at $t = 150s$, then it remained almost constant. This pattern of energy usage underscores the SNN's efficiency, particularly in the latter stages when control inputs stabilize and all LIF neurons are in standby mode. Note that the energy consumption analysis is based on the Intel's Loihi chip, a benchmark for energy-efficient neuromorphic computing.

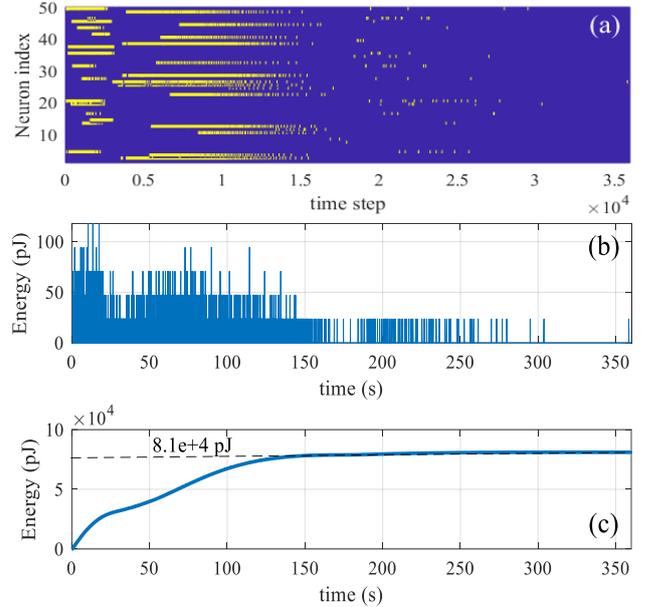

**Figure 11**. (a) Spiking activities in SNN, (b) Energy consumed per time step, and (c) Accumulative energy consumption for satellite rendezvous without obstacle.

### 4.3 Satellite rendezvous with static obstacle

To further evaluate the proposed framework's functionality and energy-efficiency in navigating around obstacles, here, the rendezvous problem has been revisited with the addition of a static obstacle. Figure 12 reveals the obtained rendezvous trajectory in the presence of a static obstacle. As it has been shown in the presented figure the interaction between the plant and the obstacle starts at $t = 22.5s$ and it ends at almost $t = 62s$. The comparison of the trajectory achieved by the system with the anticipated path outlined by cloud data confirms the effective performance of the proposed SNN in executing the assigned task. It is important to note that in practical satellite scenarios, the obstacles encountered could range from space debris to other uncontrolled objects in outer space.

Figure 13 illustrates the outcomes of controlled states for the static obstacle scenario. The figure clearly shows that, upon encountering the obstacle, there was a noticeable change in the state trajectory. However, the system, under the guidance of the SNN, adeptly followed the expected state trajectory as computed on the cloud. This adaptation highlights the SNN's capacity to effectively deviate from the initial, straightforward trajectory depicted in Figure 8, ensuring successful navigation around the obstacle.

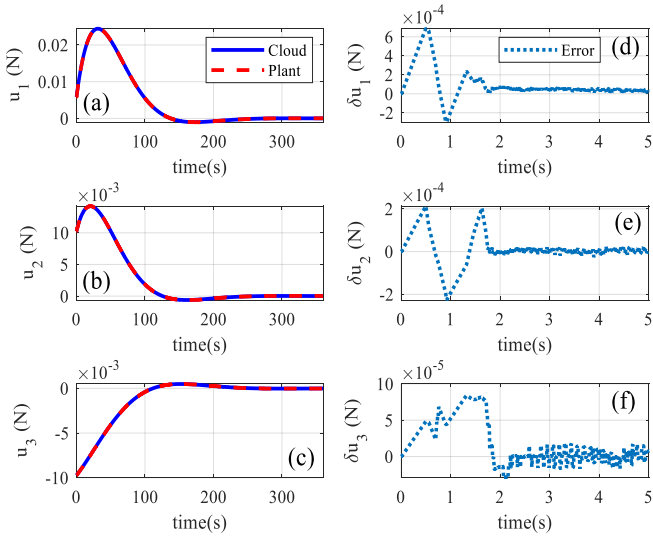

**Figure 10**. Comparation of the satellite system's control signals against cloud-provided signals in a satellite rendezvous without obstacles. Subplots (a) to (c) depict the control input vectors, highlighting the alignment and accuracy of the satellite system in the cloud directives in an obstacle-free environment. Subplots (d) to (f) focus on error convergence, showcasing the system's precision and ability to closely match the cloud's control signals, illustrating its reliability and efficiency in a clear trajectory scenario.

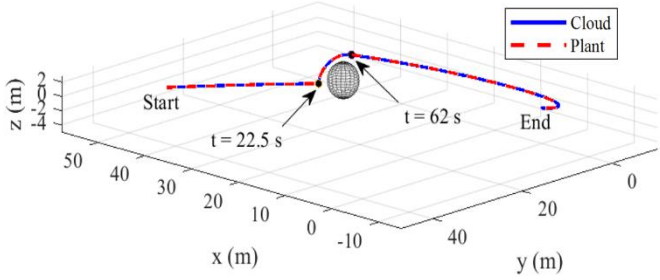

**Figure 12**. Comparison of obtained trajectory in 3D space for the plant under the SNN control with the expected trajectory provided by cloud for satellite rendezvous with a static obstacle. Note: The supplementary video provides a visual demonstration of the satellite's navigation for static obstacle avoidance with respect to spiking activities and energy consumption in Figure15.

Figures. 14(a), 14(b) and 14(c) display a comparative analysis between the control input vector elements calculated in the cloud and their equivalent outputs from the implemented SNN. The results indicate that, in scenarios involving a static obstacle, our framework effectively tracks the cloud-provided control inputs and adeptly executes the obstacle avoidance maneuver. The shaded regions in these figures denote the time intervals during which the system navigates around the obstacle. Specifically, Figure 14(a) and Figure 14(b) show that as the system approaches the obstacle, it senses its presence, and the controller correspondingly reduces the actuation levels for $u_1$, and $u_2$ from 0.022 N to approximately 0.005 N and from 0.014 N to 0.004 N, respectively. This adjustment signifies a deceleration in the x and y axes of the 3D space while the controller re-accelerates the system within the post-interaction zone. Conversely, Figure 14(c) demonstrates that in the interaction zone, the controller increases $u_3$, indicating an acceleration in the z-axis.

Moreover, Figures 14(d), 14(e) and 14(f) exhibit the time-history of tracking error for the SNN in aligning with the control signals provided by the cloud during the first 5 seconds of the simulation, as the critical stages of the control signals. Figure 14(a) shows that the error corresponding to $u_1$, nearly converges to zero by $t = 2.2s$. Similarly, Figures 14(e) and 14(f) indicate that the errors for $u_2$ and $u_3$ converge to zero around $t = 2.5s$. These observations suggest that the SNN successfully accomplishes the obstacle avoidance task with an acceptable convergence time to the desired control signal.

Figure 15(a) displays the spiking pattern of the SNN in response to the obstacle encounter scenario. The spiking pattern clearly changes as the system enters the obstacle interaction zone, reflecting SNN's adaptation to the sudden shifts in control inputs provided by the cloud. Upon exiting the interaction area, the SNN reverts to a spiking pattern similar to that observed in the obstacle-free scenario, as shown in Figure 11(a). This behavior underscores the event-driven nature of SNNs, which are capable of adjusting their neural activity in response to new conditions or significant changes in tracking errors. They achieve this with a minimal number of spikes, thereby enhancing computational efficiency. In quantitative terms, the SNN accomplished the task using 4,382 spikes, which constitutes only 0.24% of the total possible spikes. This spike count is higher in the presence of the obstacle, indicating that the network required more spikes to effectively perform the necessary computations under these more challenging conditions. This result reaffirms the adaptability and efficiency of SNNs in dynamically changing environments.

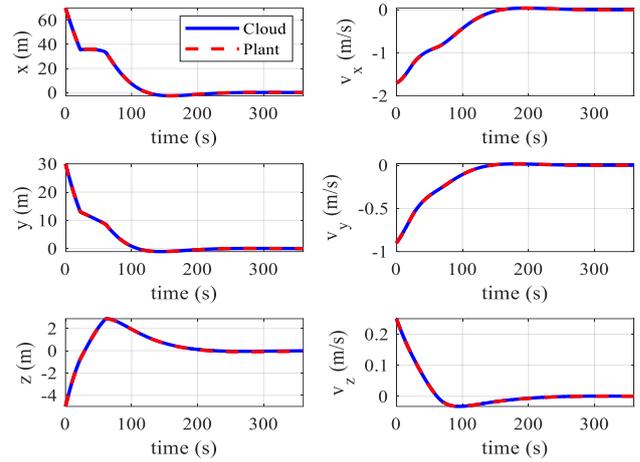

**Figure 13**. Time histories of the controlled states from the plant versus cloud results for satellite rendezvous with a static obstacle.

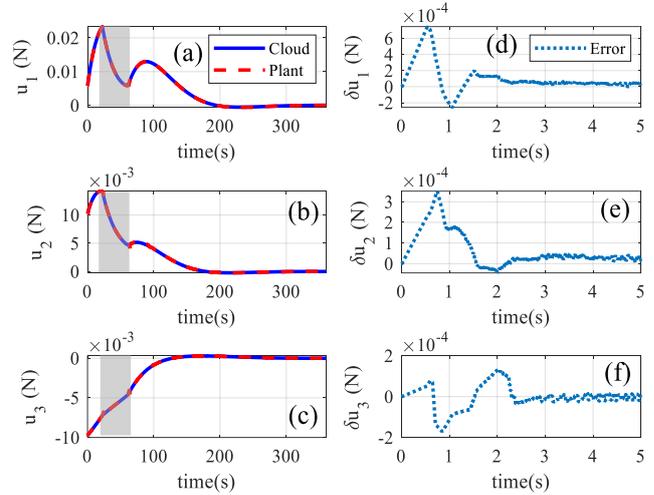

**Figure 14**. Time histories and error analysis in satellite rendezvous with a static obstacle, highlighting the comparative analysis of the control signals reproduced by the system and those generated by the cloud. (a)-(c) Clear comparison of control input vectors, demonstrating the system's response to cloud directives in the presence of a static obstacle. (d)-(f) Error convergence, emphasizing the precision and adaptability of the system in accurately following the control signals, underscoring its effectiveness in navigating around a static obstacle during the rendezvous maneuver.

Figure 15(b) illustrates the energy consumption of the SNN associated with spike emissions over time. The graph indicates a notable increase in energy usage as the system interacts with the obstacle, peaking slightly above 200 pJ. The energy consumption then decreases towards the end of the interaction zone but rises again to approximately 120 pJ shortly after exiting this area. Subsequently, it demonstrates fluctuations between 25 to 50 pJ per second until $t = 200s$, where it stabilizes. This pattern suggests that the SNN's energy consumption increases in response to environmental changes, such as the presence of an obstacle, or alterations in its operational conditions. This increase is attributable to the event-driven nature of SNNs, which dynamically adjust their processing to handle new situations.

Additionally, Figure 15(c), which presents the cumulative energy consumption of the SNN, supports these observations. It highlights a change in the slope of the energy graph at the start and end of the interaction zone. The total energy usage rises to $8.1 \times 10^4$ pJ by $t = 200s$ and then remains relatively stable, with no significant fluctuations for the remainder of the simulation. This consistency in energy consumption past the initial period of adaptation further demonstrates the efficiency of the SNN in maintaining its operation under varying conditions.

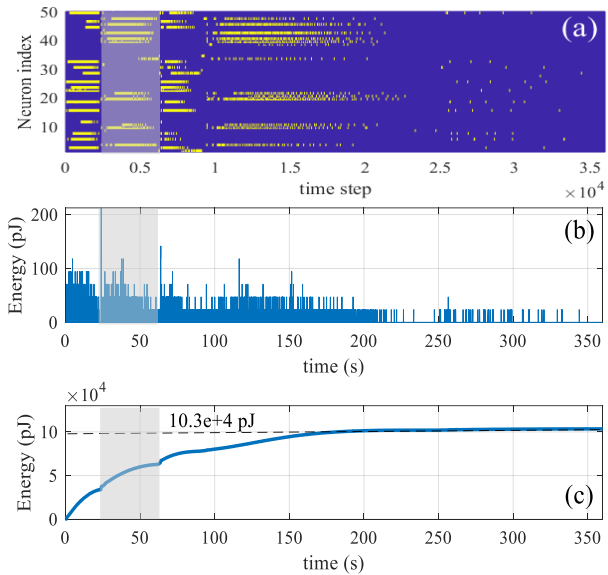

**Figure 15.** (a) Spiking pattern, (b) Energy consumed per time step, and (c) Total energy consumption for satellite rendezvous with a static obstacle.

*4.4 Satellite rendezvous with dynamic obstacle*

This section presents the results of our framework's performance in scenarios involving dynamic obstacles, building upon the previous section which focused on static obstacles. Figure 16 showcases the trajectory of a satellite rendezvous maneuver in the presence of a dynamic obstacle within a three-dimensional space. The figure indicates that the trajectory followed by the satellite, as directed by the SNN, aligns closely with the expected trajectory computed by the cloud. Our results show the satellite executed the obstacle avoidance from $t = 19s$ to $t = 27s$, effectively navigating around the dynamic obstacle. Notably, a supplementary video file is available, providing a clear visual representation of the satellite's navigation during the obstacle avoidance maneuver. A supplementary video provides a vivid representation of the navigation, enhancing understanding of the framework's real-time capabilities. It illustrates the SNN's effectiveness in executing the planned path and avoiding dynamic obstacles, demonstrating our framework's practical applicability and reliability in dynamic space navigation.

Figure 17 further elaborates on the dynamics of the satellite's interaction with the moving obstacle. It graphically represents the control inputs and trajectory adjustments as the satellite navigates the dynamic obstacle in real time. This figure provides a detailed analysis of how the satellite, under the guidance of the SNN-based controller, dynamically adjusts its path to maintain the planned trajectory while avoiding obstacles. Similar to the previous case of a static obstacle, the obtained results demonstrate the SNN's adaptability and precision in managing complex, changing scenarios, affirming its capability to modify the satellite's path. The figure highlights key moments of trajectory adjustment with regard to Figure 16 and showcases how the control inputs vary in response to the obstacle's movement.

Figures 18(a), 18(b), and 18(c) compare control inputs from the SNN on the satellite to those from the cloud, showing similar performance to static obstacle scenarios with SNN's signals closely following cloud's. A notable difference in this dynamic obstacle scenario is the extended duration of interaction between the satellite and the moving obstacle, starting at $t = 19s$ and continuing until $t = 72s$. This prolongs the obstacle avoidance maneuver by approximately 13.5 seconds compared to the static obstacle case. Figures 18(d), 18(e), and 18(f) present the tracking errors for the control inputs derived from the SNN relative to the cloud's signals. The results show that tracking convergence occurred in less than 2 seconds, indicating satisfactory performance.

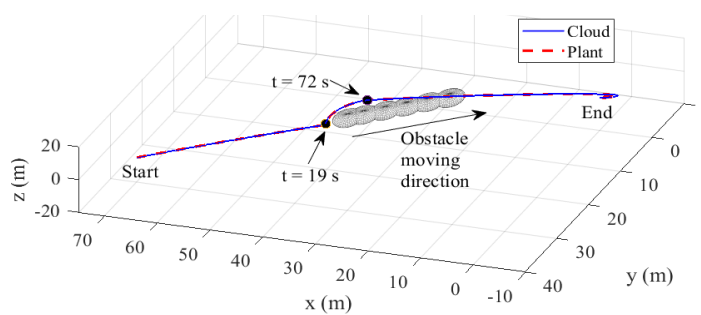

**Figure 16**. Comparison of obtained trajectory in 3D space for the plant under the SNN control with the expected trajectory provided by cloud for satellite rendezvous with a dynamic obstacle. Note: The supplementary video provides a visual demonstration of the satellite's navigation for dynamic obstacle avoidance with respect to spiking activities and energy consumption in Figure 19.

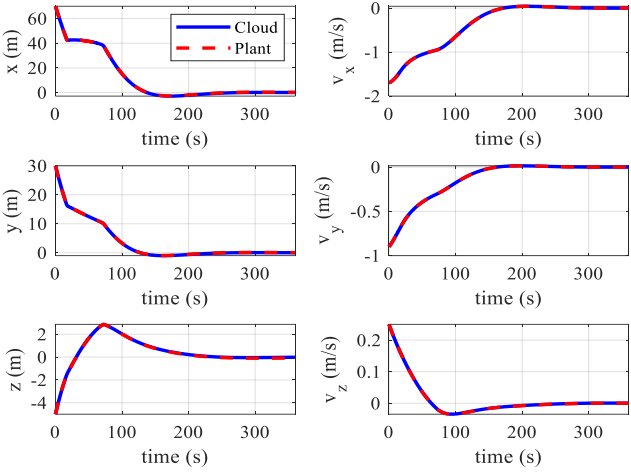

**Figure 17.** Time histories of obtained results for the controlled states from the plant in comparison with cloud results for satellite rendezvous with dynamic obstacle

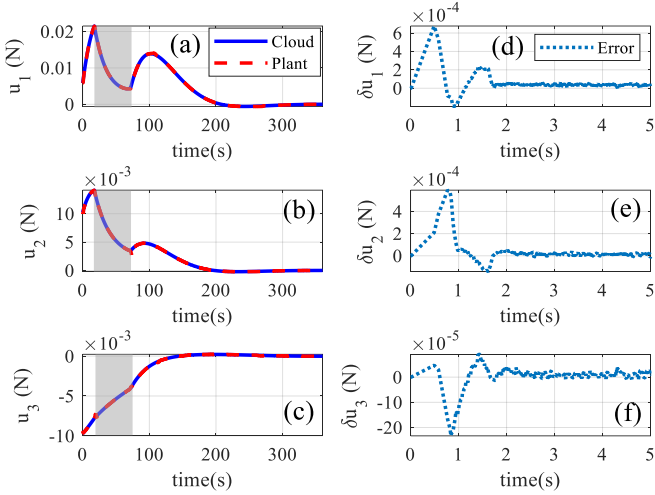

**Figure 18.** Comparative analysis and error convergence in satellite rendezvous with a dynamic obstacle. This graph presents the time histories of the reproduced control signals versus those provided by the cloud. Subplots (a) to (c) depict the comparison of control input vectors, illustrating the correlation and responsiveness of the SNN-based system to the cloud-generated directives. Subplots (d) to (f) focus on error convergence, showcasing the precision and accuracy of the SNN in tracking the provided control signals, and highlighting the efficacy of the system in real-time adjustment and navigation in the presence of a dynamic obstacle.

Figure 19(a) illustrates the spiking pattern of the SNN in the dynamic obstacle scenario. The spiking pattern during the interaction phase is highlighted to facilitate analysis. Compared to the static obstacle scenario in Figure 15(a), there's an increase in interaction length and a shift in spiking activity, demonstrating the SNN's adaptability to changing conditions, akin to biological neural responses. In this scenario, the SNN utilized 4704 spikes, about 0.26% of its potential spiking capacity. Specifically, during the interaction phase with the dynamic obstacle, SNN used 126 spikes, 17 more than in the static obstacle scenario, resulting in an additional energy consumption of 401.2 pJ on Intel's Loihi chip. Figure 19(b) shows the energy consumption per time step of the SNN in the dynamic obstacle scenario. Similar to Figure 15(b), there is a significant rise in energy use to 170 pJ at $t = 19$s, when the satellite enters the interaction zone, followed by a decrease until the end of the interaction phase. After exiting the interaction zone, the energy consumption spikes again as the SNN adjusts to the new conditions.

Figure 19(c) depicts the cumulative energy consumption, reaffirming the observations from previous analyses. The graph's slope increases at the start and end of the interaction period, indicative of heightened energy use, followed by a steady decline. Around $t = 230$s, the slope stabilizes, reflecting a consistent spiking pattern and convergence of energy consumption at approximately $11.1 \times 10^4$ pJ.

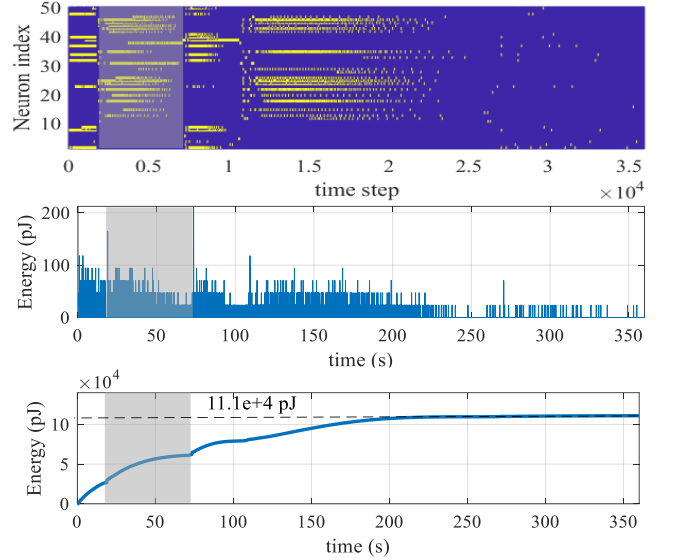

**Figure 19.** (a) Spiking pattern, (b) Energy consumed per time step, and (c) Total energy consumption for satellite rendezvous with a dynamic obstacle.

## 5 Conclusion

Addressing the challenge of resource-intensive control systems, our research presents an innovative approach to dynamic system control using Spiking Neural Networks (SNNs) for energy and computational efficiency. Leveraging cloud-edge-based learning, SNNs accurately replicate control signals for autonomous plant operation, reducing the need for constant plant-cloud communication. The architecture demonstrates a 96% reduction in tracking error with enhanced neuron count and adapts computational resources effectively to varying complexities. Notably, energy consumption increases only moderately under challenging conditions. This study underscores the potential of SNNs in real-time applications, offering a pathway for efficient and scalable control systems in robotics and beyond.

## 6 Conflict of Interest

The authors declare no conflicts of interest or relationships that could affect the study's results.

# References

[1] C. D. Schuman, S. R. Kulkarni, M. Parsa, J. P. Mitchell, P. Date and B. Kay, "Opportunities for neuromorphic computing algorithms and applications," *Nature Computational Science,* vol. 2, no. 1, pp. 10-19, 2022.

[2] M. S. Mahmoud, "Cloud-based control systems: Basics and beyond," in *Journal of Physics: Conference Series*, (Vol. 1334, No. 1, pp. 012006), 2019.

[3] J. Schlechtendahl, F. Kretschmer, Z. Sang, A. Lechler and X. Xu, "Extended study of network capability for cloud based control systems," *Robotics and Computer-Integrated Manufacturing,* vol. 43, pp. 89-95, 2017.

[4] W. Maass, "Networks of spiking neurons: the third generation of neural network models," *Neural networks,* vol. 10, no. 9, pp. 1659-71, 1997.

[5] G. Tang, Biologically Inspired Spiking Neural Networks for Energy-Efficient Robot Learning and Control, (Doctoral dissertation, Rutgers The State University of New Jersey, School of Graduate Studies), 2022.

[6] J. K. Echraghian, M. Ward, E. O. Neftci, X. Wang, G. Lenz, G. Dwivedi, M. Bennamoun, D. S. Jeong and W. D. Lu, "Training spiking neural networks using lessons from deep learning," in *Proceedings of the IEEE*, 2023.

[7] K. Yamazaki, V. K. Vo-Ho, D. Bulsara and N. Le, "Spiking neural networks and their applications: A Review," *Brain Sciences,* vol. 12, no. 7, p. 863, 2022.

[8] R. legenstein, D. Pecevski and W. Maass, "A learning theory for reward-modulated spike-timing-dependent plasticity with application to biofeedback," *PLoS computational biology,* vol. 4, no. 10, p. e1000180, 2008.

[9] X. Wang, X. Lin and X. Dang, "Supervised learning in spiking neural networks: A review of algorithms and evaluations," *Neural Networks,* vol. 125, pp. 258-280, 2020.

[10] F. ponulak and A. Kasinski, "Supervised learning in spiking neural networks with ReSuMe: sequence learning, classification, and spike shifting," *Neural computation,* vol. 22, no. 2, pp. 467-510, 2010.

[11] T. DeWolf, T. C. Stewart, J. J. Slotine and C. Eliasmith, "A spiking neural model of adaptive arm control," *Proceedings of the Royal Society B: Biological Sciences,* vol. 283, no. 1843, pp. 2016-34, 2016.

[12] A. Bougains and M. Shanahan, "Training a spiking neural network to control a 4-dof robotic arm based on spike timing-dependent plasticity," in *The 2010 International Joint Conference on Neural Networks (IJCNN)*, (pp. 1-8), 2010.

[13] A. Alemi, C. Machens, S. Deneve and J. J. Slotine, "Learning nonlinear dynamics in efficient, balanced spiking networks using local plasticity rules," in *Proceedings of the AAAI conference on artificial intelligence*, (Vol. 32, No. 1), 2018.

[14] M. Boerlin, C. K. Mchanes and S. Deneve, "Predictive coding of dynamical variables in balanced spiking networks," *PLoS computational biology,* vol. 9, no. 11, pp. e10032-58, 2013.

[15] F. S. Slijkhuis, S. W. Keemink and P. Lanillos, "Closed-form control with spike coding networks," *arXiv preprint arXiv:2212.12887,* 2022.

[16] R. Ahmadvand, S. S. Sharif and Y. M. Banad, "Enhancing Energy Efficiency and Reliability in Autonomous Systems Estimation using Neuromorphic Approach," in *arXiv preprint arXiv:2307.07963*, 2023.

[17] R. Ahmadvand, S. S. Sharif and Y. M. Banad, "Neuromorphic Robust Framework for Concurrent Estimation and Control in Dynamical Systems using Spiking Neural Networks," *arXiv preprint arXiv:2310.03873,* 2023.

[18] A. Guez and J. Selinsky, "Neurocontroller design via supervised and unsupervised learning," *Journal of Intelligent and Robotic Systems,* vol. 2, pp. 307-335, 1989.

[19] E. Okyere, A. Bousbaine, G. T. Poyi, A. K. Joseph and J. M. Andrade, "LQR controller design for quad-rotor helicopters," *The Journal of Engineering,* vol. 2019, no. 17, pp. 4003-07, 2019.

[20] A. Flores-Abad, O. Ma, K. Pham and S. Ulrich, "A review of space robotics technologies for on-orbit servicing," *Progress in aerospace sciences,* vol. 68, pp. 1-26, 2014.

[21] F. Branz, L. Savioli, A. Francesconi, F. Sansone, J. Krahn and C. Menon, "Soft docking system for capture of irregularly shaped, uncontrolled space objects," in *6th European Conference on Space Debris, ESA/ESOC, Darmstadt, Germany*, 2013.

[22] G. Arantes and L. S. Martins-Filho, "Guidance and control of position and attitude for rendezvous and dock/berthing with a noncooperative/target spacecraft," *Mathematical Problems in Engineering,* 2014.

[23] M. Kiani and R. Ahmadvand, "The Strong Tracking Innovation Filter," *IEEE Transactions on Aerospace and Electronic Systems,* vol. 58, no. 4, pp. 3261-70, 2022.

[24] M. Davies, N. Srinivasa, T. H. Lin, G. Chinya, Y. Cao, S. H. Choday, G. Dimou, P. Joshi, N. Imam, S. Jain and Y. Liao, "Loihi: A neuromorphic manycore processor with on-chip learning," *Ieee Micro,* vol. 38, no. 1, pp. 82-99, 2018.